# Fiber spectrum analyzer based on planar waveguide array aligned to a camera without lens

Xinhong Jiang[a,b], Zhifang Yang[c], Lin Wu[a,b], Zhangqi Dang[a,b], Zhenming Ding[a,b], Zexu Liu[a,b], Qing Chang[d], and Ziyang Zhang[a,b,*]

[a] *Institute of Advanced Technology, Westlake Institute for Advanced Study, 18 Shilongshan Road, Hangzhou 310024, Zhejiang Province, China.*
[b] *Laboratory of Photonic Integration, School of Engineering, Westlake University, 18 Shilongshan Road, Hangzhou 310024, Zhejiang Province, China.*
[c] *College of Electronic Engineering, Heilongjiang University, Harbin 150080, China.*
[d] *College of Media Engineering, Communication University of Zhejiang, Hangzhou 310018, China.*
*\*Corresponding author:* zhangziyang@westlake.edu.cn

**We propose and experimentally demonstrate a fiber spectrum analyzer based on a planar waveguide chip butt-coupled with an input fiber and aligned to a standard camera without any free-space optical elements. The chip consists of a single-mode waveguide to connect with the fiber, a beam broadening area, and a waveguide array in which the lengths of the waveguides are designed for both wavelength separation and beam focusing. The facet of the chip is diced open so that the outputs of the array form a near-field emitter. The far field are calculated by the Rayleigh-Sommerfeld diffraction integral. We show that the chip can provide a focal depth on the millimeter scale, allowing relaxed alignment to the camera without any fine-positioning stage. Two devices with 120 and 220 waveguides are fabricated on the polymer waveguide platform. The measured spectral width are 0.63 nm and 0.42 nm, respectively. This simple and practical approach may lead to the development of a spectrum analyzer for fiber that is easily mountable to any commercial camera, thereby avoiding the complication for customized detectors as well as electronic circuits afterwards.**

## 1. Introduction

Spectroscopy is a fundamental tool in the research fields of physics, chemistry, astronomy, etc. By analyzing the electromagnetic spectra, the composition and structures of matter can be investigated from atomic to astronomical scales [1–4]. For many spectroscopic applications, a miniature wavelength analyzing device is desired for compact and potentially low-cost system integration.

In recent years, arrayed waveguide gratings (AWGs) have found wide applications in spectroscopic studies beyond wavelength-division-multiplexing technology in optical communication [5,6]. However, AWG itself is a passive device and must be integrated with an array of photodiodes (PDs) to convert the optical signals at different wavelengths into physically separated electronic signals (photocurrent). The integrated assembly has been used to demodulate multi-wavelength signals from a fiber Bragg grating sensor [7].

The integration of AWG and PDs can be done either monolithically or in a hybrid manner. Silica and polymer-based AWGs adopt mostly the hybrid integration approach. Planar PD arrays can be placed on top of the chip via an integrated 45-degree mirror [8,9]. Since the waveguide structure is on the micrometer scale, the alignment accuracy must also be on the same order to ensure an efficient optical coupling and low crosstalk among the PDs. Monolithic integration takes place mostly on silicon and InP platforms, where Ge and InP PDs can be fabricated on the same device, respectively [10,11]. However, the subsequent electronic amplifiers (TIAs) and processing circuits still need to be customized to the layout of the PDs. The integration of the customized electronic units inevitably increases the cost of the system unless the demand scales up to large-scale production. The non-standard circuit board may largely increase the footprint of the spectrum analyzer, resulting in an overall bulky device, despite the compact size of the AWG itself, thereby limiting the flexible and low-cost application of the AWG-based solution.

There have been attempts to build a spectrometer, or "spectrograph" in astronomy terms, using AWG and an external camera to collect and analyze the dispersed light [12–15]. In this configuration, the output waveguides along the output facet of the second slab of an AWG are cut away, allowing light to emit from the slab. There are two major problems. Firstly, since light is already focused on the output facet of the second slab within the chip, further propagation in free space results in beam diverging. Therefore, a free-space lens is needed to refocus light on the camera. Though AWG itself is compact and lightweight, the added elements still increases the system size and complexity. Secondly, the chip needs to be diced precisely at the output facet of the second slab. The curvature of the facet causes aberration for off-axis beams, which decreases the resolution at the off-axis wavelengths.

In this work, we come up with a new strategy and demonstrate a planar waveguide chip, named waveguide spectral lens (WSL), which can fulfill both wavelength separation and beam focusing functions. The output waveguides are straight to the dicing facet and dispersed light can be focused onto a commercial camera placed at arbitrary distance, all without free-space optical elements. Laterally, the focal plane can match the sensor size of the camera, allowing a full free spectral range (FSR) to be covered. The focal depth can be enlarged to a few millimeters, allowing relaxed alignment between the chip and the camera. By changing the number of waveguides, different spectral resolutions can be achieved under similar FSRs. The devices are fabricated on a low-cost polymer waveguide platform as proof of concept.

## 2. Device structure and operation principle

Fig. 1 shows the function of the proposed WSL. Light from the input fiber is butt-coupled to a single mode waveguide and guided to a slab region as beam broadening area (BBA), where the beam diverges horizontally but remains confined in the vertical dimension. At the output of the BBA, a number of waveguides are placed with equal spacing along the periphery of the slab and have equal distance to the center of front facet. Thus, the broadened wavefront of the input light is collected by the waveguides with approximately the same phase. The lengths of the waveguides are varied to control the phase for the grating effect as well as the beam-focusing functions. The waveguides are arranged with equal spacing at the output facet. When emitting in free space, light with modulated phase front creates a dispersed yet focused line jointly on the camera.

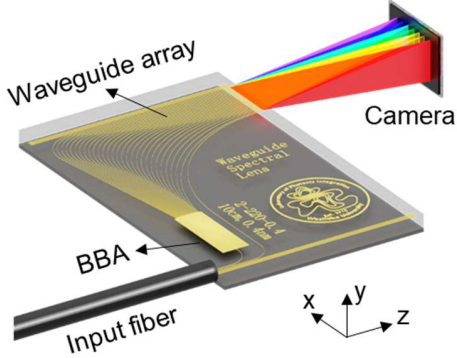

**Fig. 1.** Schematic of the proposed spectrum analyzing device for an optical fiber using a WSL and a standard camera. WSL: waveguide spectral lens. BBA: beam broadening area.

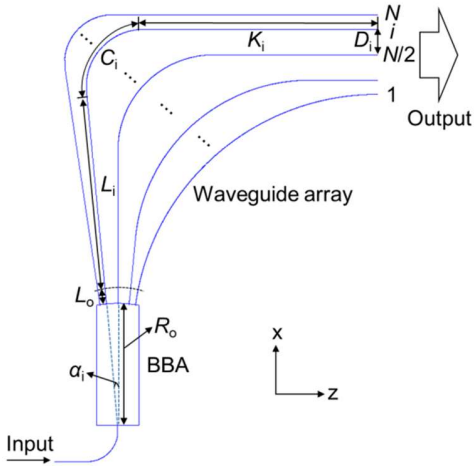

**Fig. 2.** Device layout and waveguide arrangement of the WSL.

Fig. 2 shows the device layout and design. The length of the $i$-th waveguide is defined as $l_i$. The length difference of the $i$-th waveguide to the first waveguide is designed as:

$$l_i - l_1 = P_i - P_1 + (i-1)\Delta L, (i=1,...,N), \quad (1)$$

$$P_i = (f - \sqrt{f^2 + D_i^2})/n_{\text{eff}}, \quad (2)$$

$$\Delta L = 1/\left(n_{\text{eff}}/\lambda_{\text{start}} - n_{\text{eff}}/\lambda_{\text{stop}}\right), \quad (3)$$

$$\text{FSR} = \lambda_{\text{stop}} - \lambda_{\text{start}} \approx \lambda^2/(n_{\text{eff}}\Delta L), \quad (4)$$

where $P_i$ is the required relative length in the $i$-th waveguide for the beam focusing function [16]. $\Delta L$ is the uniform length difference of any two adjacent waveguides for wavelength separation. $N$ is the total number of the waveguides. $f$ is the focal length, i.e., the chosen distance in free space where the camera should be placed. $D_i$ is the distance between the $i$-th waveguide output facet and the center of the whole output facet. $n_{\text{eff}}$ is the effective index of the waveguide. $\lambda_{\text{start}}$ and $\lambda_{\text{stop}}$ are the start and stop wavelengths within one FSR.

In Fig. 2, the $i$-th waveguide path in the array can be broken into a taper segment Lo, in order to vary the waveguide width smoothly to that of a single-mode waveguide, connected with a straight segment Li, an arc segment Ci, and the output straight segment Ki. The two straight segments are connected by the arc segment. The lengths of the four segments are $L_o$, $L_i$, $C_i$, and $K_i$, respectively. The corresponding length of the $i$-th waveguide is $l_i = L_o + L_i + C_i + K_i$. These lengths should be designed to satisfy the length difference requirement in Eq. (1). The radius and angle of the arc segment Ci are $R_i$ and $\pi/2-\alpha_i$, respectively. $\alpha_i$ is the angle between the central line of the taper and the $x$ axis. $\alpha_i$ is defined as negative and positive values for waveguides connecting with the left and right parts of the BBA, respectively. $R_o$ is the radius of the BBA. $D_o$ is the length of the arc connecting adjacent waveguides at the output facet of the BBA.

The deflection angle $\theta$ of the far field of the WSL can be calculated as [17]

$$\theta = \arcsin(\Delta\varphi/(kd)), \quad (5)$$

where $\Delta\varphi = n_{\text{eff}}k\Delta L$ is the uniform phase difference for the grating effect, with $k$ denoting the wavenumber. $d$ is the waveguide spacing at the output facet of the waveguide array. To calculate $\theta$, $\Delta\varphi$ should be normalized to the range from $-\pi$ to $\pi$.

The far-field intensity distribution of the WSL is simulated by Rayleigh-Sommerfeld diffraction integral [18]. Firstly, the TE mode profile of the input waveguide is simulated using the eigenmode solver in Lumerical. The waveguide cross section is 3 μm by 3 μm. The cladding and core indexes are 1.45 and 1.48, respectively. Then the amplitudes and phases of the input fields of the waveguides in the array are calculated by the diffraction integral of the TE mode profile of the input waveguide by setting the refractive index of the medium space as the effective index of the BBA [18]. The phase shifts in each waveguide can be calculated based on Eq. (1). Finally, the far-field at a fixed distance is calculated by diffraction integral of the output field of the waveguide array. The lateral spread-out of the far field in $x$ axis of the wavelengths within one FSR is defined as $\Delta x$, which should be narrower than the width of the camera sensor. $\Delta x$ can be approximately calculated as

$$\Delta x = 2f \tan(\text{FOV}/2), \quad (6)$$

where FOV = $2\arcsin(\lambda/(2d))$ is the field of view of the far field [19].

**Table 1**
Design and performance parameters of the WSLs

| Parameter | WSL1 | WSL2 |
|---|---|---|
| $N$ | 120 | 220 |
| FSR (nm) | 64 | 62 |
| $\Delta L$ (μm) | 25.7 | 26.5 |
| $f$ (cm) | 10 | 10 |
| $d$ (μm) | 33 | 18 |
| $R_o$ (μm) | 6000 | 6000 |
| $D_o$ (μm) | 8 | 8 |
| Calculated $\Delta x$ (μm) | 4654.8 | 8611.1 |
| Simulated focal depth (mm) | 4.59 | 4.32 |
| Simulated FWHM (nm) | 0.50 | 0.29 |
| Measured FWHM (nm) | 0.63 | 0.42 |
| Measured FSR (nm) | 62.12 | 59.45 |
| Footprint (mm$^2$) | 15.2 × 6.8 | 15.0 × 5.2 |

One key parameter of a spectral analyzer is the spectral resolution, which can be characterized as the full width at half maximum (FWHM) in the spectrum plot. Different FWHMs of the far-field distribution can be obtained by changing the number of waveguides [17]. To investigate the influence of the number of waveguides, two WSLs with different numbers of waveguides and similar FSRs are designed. The numbers of waveguides of the WSLs are 120 and 220, respectively. The WSLs are named WSL1 and WSL2, respectively. The simulation parameters FSR, $\Delta L$, $f$, $d$, $R_o$, and $D_o$ are listed in Table 1. The $\Delta x$ calculated by Eq. (6) are 4654.8 μm and 8611.1 μm for the 120 and 220-waveguide designs, respectively. As the camera is equipped with a sensor of 640 × 512 pixels (20 μm pixel pitch), the spectral lines of a whole FSR can be captured. The simulated far-field intensity distributions of the WSLs at 1550 nm are shown in Fig. 3. The simulated FWHMs are 0.50 nm and 0.29 nm, respectively, which decrease with the increasing of the number of waveguides.

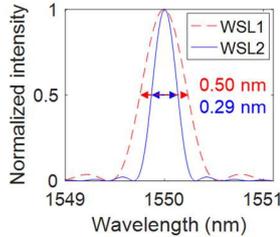

**Fig. 3** Simulated far-field intensity distributions of WSL1 and WSL2. The input wavelength is 1550 nm.

The focal depth of the WSL is another key parameter as a large focal depth can relax the alignment between the chip and the camera. Fig. 4(a) shows a simplified model for the illustration of the focal depth of a WSL. The FWHM of the far field ($\Delta w$) is proportional to $|\Delta z|(N-1)d/f$, where $|\Delta z|$ is the distance between the detector plane and the focal plane. ($N$–1)$d$ is the output facet width of the waveguide array. To study the focal depth numerically, we move the detector plane before and after the focal plane and calculate the modified far-field profiles. The threshold is set to 10% widening of the optimal FWHM, i.e, the spectral linewidth at the focal plane. The results are displayed in Fig. 4(b) and (c). The $x$ positions of the far-field peaks for WSL1 and WSL2 at 1550 nm are positive and negative values, respectively, according to Eq. (5). For the design with 120 waveguides, the camera sensor can be placed at –2.45 to +2.14 mm away from the focal plane, corresponding to a focus depth of 4.59 mm. The focal depth for the WSL with 220 waveguides is 4.32 mm. Since the output facet widths of the two WSLs are close, they have similar focal depths according to the analysis of Fig. 4(a).

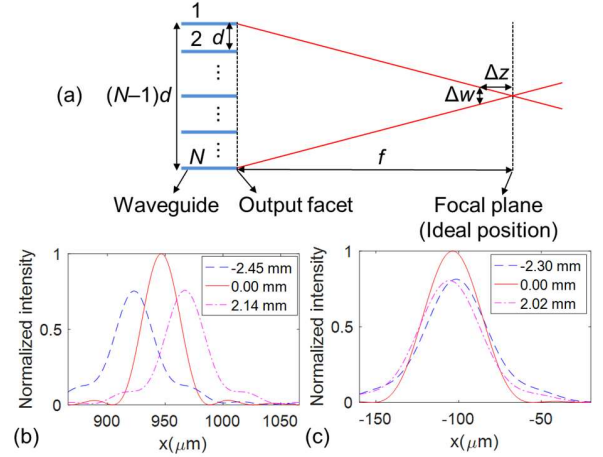

**Fig. 4.** (a) Simplified model for illustration of the focal depth of a WSL. The ideal position of the camera sensor is at the focal plane. (b,c) Simulated far-field intensity distributions at 1550 nm of the (b) 120 and (c) 220-waveguide WSLs with different camera positions and an FWHM broadening threshold of 10%. The dashed and dash-dotted lines show the intensity distributions when the camera sensor is not placed at the focal plane.

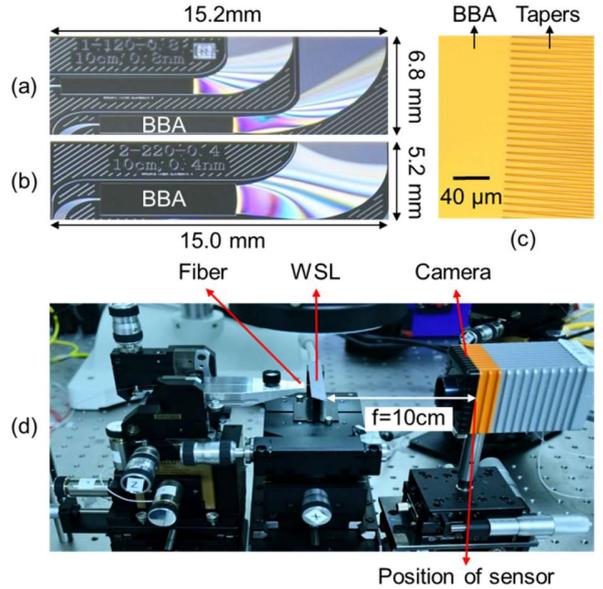

**Fig. 5.** (a)(b) Photos of the fabricated WSLs with (a) 120 and (b) 220 waveguides. (c) Photos of the output facet of the BBA and the tapers. (d) Photo of the WSL testing system.

## 3. Device fabrication and experimental results

The designed devices are fabricated on a polymer platform. The fabrication process is similar to that of [20]. The chip is diced with a standard sawing equipment without facet polishing. The photos of the fabricated devices are shown in Fig. 5(a) and (b). The design and performance parameters are summarized in Table 1.

Fig. 5(c) shows the photo of the WSL testing system. TE-polarized light in the fiber is butt-coupled to the input waveguide of the WSL. The far field of the WSL is captured by an infrared camera without any free-space lens. The camera is placed near the focal plane of the WSL and adjusted by a coarse mechanical positioning system to reach the focal length of $f$ = 10 cm.

It is worth noting that the fiber-chip attachment is a standard technology widely used in photonic modules including AWGs. It can be done all passively via on-chip U-grooves [8] and V-grooves [21]. In our experiment, the placement of the camera is firstly done by measuring the distance to the chip facet by a ruler and the spectral lines are already well focused. The position is then adjusted slightly via the mechanical knob to achieve the sharpest line for characterization. The relaxed alignment procedure is expected from the calculations shown in Fig. 4.

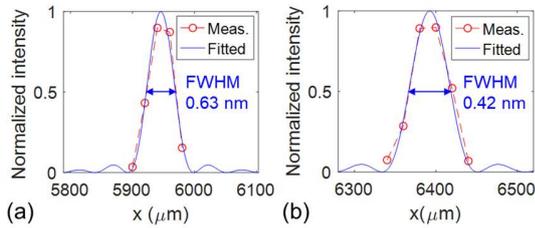

**Fig. 6.** Normalized measured and fitted far-field intensity distributions of the (a) 120 and (b) 220-waveguide WSLs.

To characterize the WSLs, a tunable laser is connected as input. The spectral lines are recorded by the camera and the photo counts from the pixels in each column are summed up. Then the peak position of each spectral line is calculated by the centroid algorithm [22]. The far-field intensity distributions are fitted by multiple-slit diffractions [23], as shown in Fig. 6. Based on the slopes of the wavelength calibration lines discussed later in this section and the fitted intensity distributions, the FWHMs of the fitted intensity distributions at 1550 nm are found to be 0.63 nm and 0.42 nm for the WSLs with 120 and 220 waveguides, respectively.

The difference of FWHMs between the simulation and experiment can be attributed to inaccuracy in setting the camera position/focal length, variation of the waveguide effective index due to fabrication tolerance (random phase errors), transmission loss imbalance of the waveguides, as well as the discretization of the far-field intensity distribution by the camera pixels.

Fig. 7(a) and (b) show the captured photos of the spectral lines taken from the 120 and 220-waveguide WSLs, respectively. The FSRs of the 120 and 220-waveguide WSLs are 62.12 nm and 59.45 nm, respectively. It can be seen that the far fields are well focused in the horizontal direction due to the applied phase modulation of the WSLs. As there is no phase control in the vertical direction, the far field diverges, similar to a conventional single-mode waveguide.

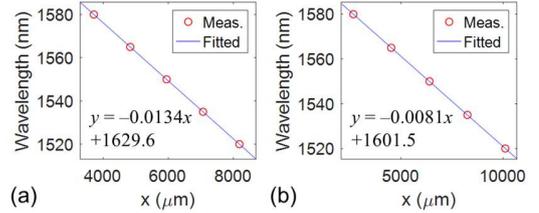

**Fig. 8.** Wavelength calibration results of the (a) 120 and (b) 220-waveguide WSLs.

To determine the wavelength of any spectral line captured by the camera, the WSL should be first calibrated using a known light source. The peak positions of five spectral lines are calculated by the centroid algorithm, then the relation between the wavelengths and the horizontal locations of the intensity peaks are linearly fitted. The results are shown in Fig. 8, indicating a good agreement between calculation and experiment.

### 4. Loss assessment

The optical loss of the device comes from the following parts: 1) coupling loss with the fiber, 2) truncation loss at the output facet of the BBA, 3) propagation loss within the WSL, 4) diffraction loss due to multiple peaks/orders, and 5) capture loss at the camera as light is not focused in the vertical direction and the sensor provides only a limited height.

The fiber-waveguide coupling loss and the waveguide propagation loss can be obtained from a standard cut-back measurement with waveguides of different lengths, fabricated on the same wafer and with the same design. The waveguide propagation losses for the TE and TM polarizations are ~0.87 dB/cm and ~0.79 dB/cm, respectively. The fiber-waveguide coupling losses for the TE and TM polarizations are ~2.42 dB/facet and ~2.41 dB/facet, respectively. The fiber coupling loss can be reduced to below 0.25 dB/facet by designing a mode-matched taper at the input side of the waveguide [8]. The propagation loss can be further reduced by waveguide technology in other platforms, such as silica on silicon.

The truncation loss, diffraction loss, and capture loss can all be calculated based on diffraction integrals. The diffraction integral for calculating the truncation loss is along the horizontal direction from the input waveguide to the output facet of the BBA. The diffraction integrals for calculating the diffraction loss and capture loss are along the horizontal and vertical directions from the output facet of the waveguide array to the focal plane, respectively. The simulated intensity distributions of the diffraction integrals for calculating the losses of WSL1 and WSL2 are shown in Fig. 9. The light is mainly located in the 1-rad range of the far-field intensity distribution. The loss can be calculated as the ratio of the collected light intensity to the total intensity. The calculated truncation loss, diffraction loss, and capture loss are 1.6 dB, 8.9 dB, and 3.0 dB for WSL1, and 0.4 dB, 6.2 dB, and 3.0 dB for WSL2, respectively, as shown in Fig. 9.

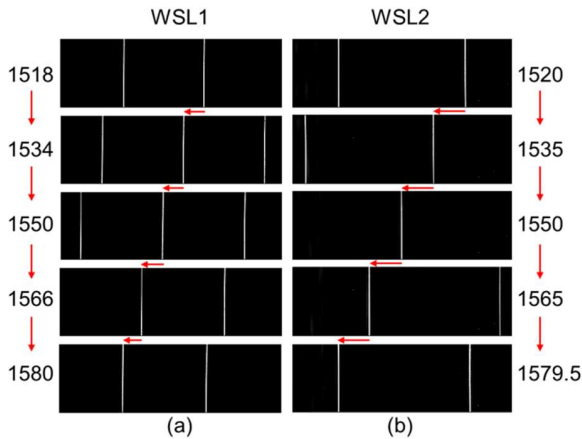

**Fig. 7.** Captured spectral lines at different wavelengths of the (a) 120 and (b) 220-waveguide WSLs. The red arrows indicate the wavelength variation and the corresponding shift of the spectral lines.

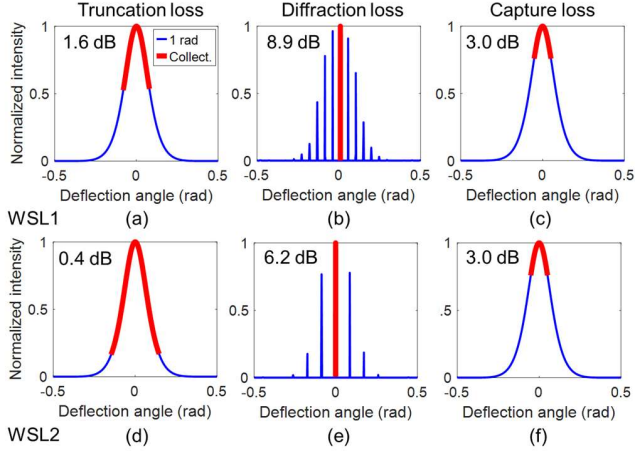

**Fig. 9.** Simulated intensity distributions at 1550 nm for calculating the truncation loss, diffraction loss, and capture loss of (a,b,c) WSL1 and (d,e,f) WSL2. The structural parameters of the two devices are listed in Table 1. The thin-blue line is the far field intensity distribution within the 1-rad range. The thick-red line is the collected light intensity. The legends for (a) to (f) are the same.

The truncation loss at the output facet of the BBA can be reduced by increasing the number of waveguides, thus more output power of the BBA can be received by the waveguide array. This is verified by the reduced truncation loss of WSL2 compared to WSL1, as shown in Fig. 9(a) and (d), since they have the same $R_o$ and $D_o$ but different numbers of waveguides.

To evaluate the chip insertion loss including fiber-chip coupling loss, propagation loss and truncation loss, we have measured the transmission at 1550 nm from the input fiber to the input waveguide and from the output fiber coupled with each output waveguide of WSL1, i.e., 120 transmission loss measurement in total. The total insertion loss sums up to 8.88 dB for the TE mode and 8.41 dB for the TM mode, respectively.

Out in free space, the relatively large diffraction loss is caused by the narrow angular spacing between adjacent diffraction orders, which increases the intensities of the side lobes, as shown in Fig. 9(b) and (e). The diffraction loss can be reduced by decreasing the output facet width of the waveguide array so that the FOV is increased. As a result, the intensities of the side lobes are decreased since the envelope of the intensity distribution is determined by the far field of a single waveguide facet [23]. For example, when the waveguide spacing at the output facet is reduced to 9 μm, both the diffraction losses of the WSLs with 120 and 220 waveguides can be reduced to 3.4 dB and 3.2 dB, respectively. The simulated intensity distributions for the WSLs when $d$ is 9 μm are shown in Fig. 10(a) and 10(b), respectively.

It is worth noting that the side lobes / higher-order diffractions of the far-field spectrum from a WSL can be neglected and do not become crosstalk in the detected signal because the higher-order signals can be blocked and only the ground-order is captured by the camera. This is fundamentally different from an optical phase array used in light detection and ranging (LiDAR) where the reflected lights are detected [24]. The reflected lights of the side lobes from objects may introduce crosstalk in the detected signal as the reflected powers of different diffraction orders can be added up by the same detector.

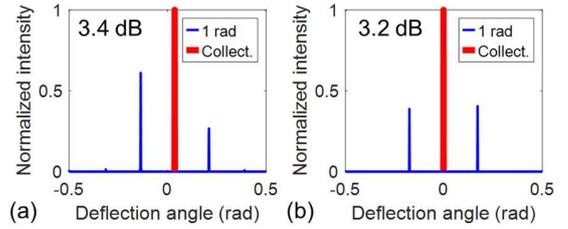

**Fig. 10.** Simulated intensity distributions at 1550 nm for calculating the diffraction loss of the WSLs with (a) 120 and (b) 220 waveguides. $d = 9$ μm.

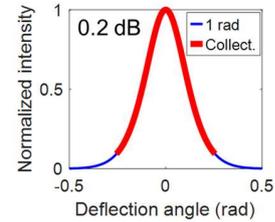

**Fig. 11.** Simulated intensity distributions at 1550 nm for calculating the capture loss of the WSLs. $f = 2$ cm.

The capture loss can be countered by placing the camera closer to the chip. Since the waveguide heights and the focal lengths of WSL1 and WSL2 are the same, they have the same capture loss. For example, when the focal length is reduced to 2 cm, most the vertical line can be captured by the camera and the capture losses of the WSLs can be reduced to 0.2 dB. The simulated intensity distributions are shown in Fig. 11. We have only designed WSL with focal length of 10 cm in this work as proof of concept. A shorter focal length is beneficial for the reduction of capture loss as well as more compact system integration.

## 5. Summary

In conclusion, we have proposed and experimentally demonstrated a fiber spectrum analyzer based on planar waveguide array aligned to a camera. The device can not only separate the wavelength spatially, but also focus them on a camera placed at a chosen distance in the system without a free-space lens. In traditional schemes, the integration of an AWG and a PD array needs customized detectors and electronic circuits, as well as alignment precision in micrometer scale. However, in this work, a commercial camera can be placed at the focal plane with low requirement on alignment precision, thanks to the large focal depth of ∼4mm. As future work, we plan to investigate the WSL device in detail, e.g., finding the limits on spectral resolution and FSR, as well as to come up with concrete mounting / integration plan to a commercial camera. Furthermore, multilayer waveguide array can be fabricated on polymer platform, allowing phase modulation also in the vertical direction and thereby enabling the possibility to develop a truly 2D dispersive device, i.e., the far field is also dispersed and focused in the vertical dimension, rendering a spectral spot map, instead of conventional spectral lines. The proposed device may find important applications in optical fiber sensing and other research fields where spectral analysis is needed.

**Funding.**
This work was supported by the National Natural Science Foundation of China (61905202).

**Declaration of Competing Interest**
The authors declare that they have no known competing financial interests or personal relationships that could have appeared to influence the work reported in this paper.

**References**
[1] R. Gutzler, M. Garg, C. R. Ast, K. Kuhnke, and K. Kern, Light–matter interaction at atomic scales, Nat. Rev. Phys. 3 (2021) 441–453.
[2] C. Gu, Z. Zuo, D. Luo, Z. Deng, Y. Liu, M. Hu, and W. Li, Passive coherent dual-comb spectroscopy based on optical-optical modulation with free running lasers, PhotoniX 1, 7 (2020).
[3] Y. Peng, C. Shi, Y. Zhu, M. Gu, and S. Zhuang, Terahertz spectroscopy in biomedical field: a review on signal-to-noise ratio improvement, PhotoniX 1, 2 (2020).
[4] P. Gatkine, S. Veilleux, and M. Dagenais, Astrophotonic spectrographs, Applied Sciences 9 (2) (2019) 290.
[5] A. V. Wijk, C. R. Doerr, Z. Ali, M. Karabiyik, and B. I. Akca, Compact ultrabroad-bandwidth cascaded arrayed waveguide gratings, Opt. Express 28 (10) (2020) 14618–14626.
[6] B. I. Akca and C. R. Doerr, Interleaved silicon nitride AWG spectrometers, IEEE Photon. Technol. Lett. 30 (1) (2019) 90–93.
[7] M. Słowikowski, A. Kazmierczak, S. Stopinski, M. Bieniek, S. Szostak, K. Matuk, L. Augustin, and R. Piramidowicz, Photonic integrated interrogator for monitoring the patient condition during MRI diagnosis, Sensors 21 (12) (2021) 4238.
[8] Z. Zhang, N. Mettbach, C. Zawadzki, J. Wang, D. Schmidt, W. Brinker, N. Grote, M. Schell and N. Keil, Polymer-based photonic toolbox: passive components, hybrid integration and polarisation control, IET Optoelectronics 5 (5) (2011) 226–232.
[9] S. Yun, Y. Han, S. Kim, J. Shin, S. Park, D. Lee, S. Lee, and Y. Baek, Compact hybrid-integrated 4 × 80-Gbps TROSA module using optical butt-coupling of DML/SI-PD and silica AWG chips, J. Lightw. Technol. 39 (8) (2021) 2468–2475.
[10] X. Luo, J. Song, X. Tu, Q. Fang, L. Jia, Y. Huang, T. Liow, M. Yu, and G. Lo, Silicon-based traveling-wave photodetector array (Si-TWPDA) with parallel optical feeding, Opt. Express 22 (17) (2014) 20020–20026.
[11] Q. Lv, Q. Han, P. Pan, H. Ye, D. Yin, and X. Yang, Monolithic integration of a InP AWG and InGaAs photodiodes on InP platform, Opt. & Laser Technol. 90 (2017) 122–127.
[12] A. Stoll, Z. Zhang, R. Haynes, and M. Roth, High-resolution arrayed-waveguide-gratings in astronomy: design and fabrication challenges, Photonics 4 (2017) 4020030.
[13] N. Cvetojevic, 1 J. S. Lawrence, S. C. Ellis, J. Bland-Hawthorn, R. Haynes, and A. Horton, Characterization and on-sky demonstration of an integrated photonic spectrograph for astronomy, Opt. Express 17 (21) (2009) 18643–18650.
[14] N. Cvetojevic, N. Jovanovic, C. Betters, J. S. Lawrence, S. C. Ellis, G. Robertson, and J. Bland-Hawthorn, First starlight spectrum captured using an integrated photonic micro-spectrograph, Astronomy & Astrophysics 544 (8) (2012) 453–478.
[15] N. Cvetojevic, N. Jovanovic, J. Lawrence, M. Withford, and J. Bland-Hawthorn, Developing arrayed waveguide grating spectrographs for multi-object astronomical spectroscopy, Opt. Express 20 (3) (2012) 2062–2072.
[16] Z. Wang, T. Li, A. Soman, D. Mao, T. Kananen, and T. Gu, On-chip wavefront shaping with dielectric metasurface, Nat. Commun. 10 (2019) 3547.
[17] K. V. Acoleyen, W. Bogaerts, J. Jágerská, N. L. Thomas, R. Houdré, and R. Baets, Off-chip beam steering with a one-dimensional optical phased array on silicon-on-insulator, Opt. Lett. 34 (9) (2009) 1477–1479.
[18] N. A. Ochoa, Alternative approach to evaluate the Rayleigh-Sommerfeld diffraction integrals using tilted spherical waves, Opt. Express 25 (10) (2017) 12008–12019.
[19] S. Chung, H. Abediasl, and H. Hashemi, A monolithically integrated large-scale optical phased array in silicon-on-insulator CMOS, IEEE J. Solid-State Circuits 53 (1) (2018) 275–296.
[20] X. Jiang, Z. Yang, Z. Liu, Z. Dang, Z. Ding, Q. Chang, and Z. Zhang, 3D integrated wavelength demultiplexer based on a square-core fiber and dual-layer arrayed waveguide gratings, Opt. Express 29 (2) (2021) 2090–2098.
[21] C. Strandman and Y. Bäcklund, Bulk silicon holding structures for mounting of optical fibers in V-grooves, J. Microelectromechanical Systems 6 (1) (1997) 35–40.
[22] X. Ma, C. Rao and H. Zheng, Error analysis of CCD-based point source centroid computation under the background light, Opt. Express 17 (10) (2009) 8525–8541.
[23] W. Xu, L. Zhou, L. Lu, and J. Chen, Aliasing-free optical phased array beam-steering with a plateau envelope, Opt. Express 27 (3) (2019) 3354–3368.
[24] C. V. Poulton, A. Yaacobi, D. B. Cole, M. J. Byrd, M. Raval, D. Vermeulen, and M. R. Watts, Coherent solid-state LIDAR with silicon photonic optical phased arrays, Opt. Lett. 42 (20) (2017) 4091–4094.